\documentclass[11pt,preprint]{aastex}

\begin{document}

\def\lapp{\ifmmode\stackrel{<}{_{\sim}}\else$\stackrel{<}{_{\sim}}$\fi}
\def\gapp{\ifmmode\stackrel{>}{_{\sim}}\else$\stackrel{>}{_{\sim}}$\fi}
\def\psr{PSR~J1811$-$1925}

\title{{\it Chandra} X-ray Observations of G11.2$-$0.3:  Implications
for Pulsar Ages}

\author{
V. M. Kaspi,\altaffilmark{1,2,3}
M. E. Roberts,\altaffilmark{1,2,4}
G. Vasisht,\altaffilmark{5}
E. V. Gotthelf,\altaffilmark{6}
M. Pivovaroff,\altaffilmark{7}
N. Kawai,\altaffilmark{8,9} }

\altaffiltext{1}{Department of Physics, Rutherford Physics Building,
McGill University, 3600 University Street, Montreal, Quebec,
H3A 2T8, Canada}

\altaffiltext{2}{Department of Physics and Center for Space Research,
Massachusetts Institute of Technology, Cambridge, MA 02139}

\altaffiltext{3}{Alfred P. Sloan Research Fellow}

\altaffiltext{4}{Quebec Merit Fellow}

\altaffiltext{5}{Jet Propulsion Laboratory, California Institute of Technology, 4800 Oak Grove Drive, Pasadena, CA 91109}

\altaffiltext{6}{Columbia University Astronomy Department, Pupin Hall, New York, 550 West 120th Street, New York, NY 10027}

\altaffiltext{7}{Therma-Wave, Inc., 1250 Reliance Way, Fremont, CA 94539}

\altaffiltext{8}{Department of Physics, Tokyo Institute of Technology, 2-12-1 Ookayama, Meguro-ku, Tokyo 152-8551, Japan}

\altaffiltext{9}{RIKEN (The Institute of Physical and Chemical Research), 2-1 Hirosawa, Wako, Saitama 351-0198, Japan}

\begin{abstract}

We present {\it Chandra X-ray Observatory} imaging observations of the
young Galactic supernova remnant G11.2$-$0.3.  The image shows that the
previously known young 65-ms X-ray pulsar is at position (J2000) RA 
18$^{\rm h}$~11$^{\rm m}$~29$^{\rm s}$.22, DEC $-19^{\circ} \; 25' \;
27.''6$, with $1\sigma$ error radius $0.''6$.  This is within 8$''$ of
the geometric center of the shell.  This provides strong confirming
evidence that the system is younger, by a factor of $\sim$12, than the
characteristic age of the pulsar.  The age discrepancy suggests that
pulsar characteristic ages can be poor age estimators for young
pulsars.  Assuming conventional spin down with constant magnetic field
and braking index, the most likely explanation for the age discrepancy
in G11.2$-$0.3 is that the pulsar was born with a spin period of
$\sim$62~ms.  The {\it Chandra} image also reveals, for the first time,
the morphology of the pulsar wind nebula.  The elongated hard-X-ray
structure can be interpreted as either a jet or a Crab-like torus seen
edge on.  This adds to the growing list of highly aspherical pulsar
wind nebulae and argues that such structures are common around young
pulsars.

\end{abstract}


\section{Introduction}
\label{sec:intro}

Determining the ages of neutron stars is important for several
reasons.  First, ages are crucial for establishing the number and birth
rate of neutron stars in the Galaxy.  This is useful for comparison
with the Galactic supernova rate, ultimately in order to establish the
fraction and types of supernovae that lead to the formation of a
neutron star.  Second, one of the few experimental constraints on the
nature of matter at very high densities comes from models of neutron
star cooling.  In order to test such models, a reliable neutron star
age is key.  The latter can help constrain the age of associated
objects, such as supernova remnants, which is important for
understanding remnant evolution and morphology.  Also, ages can
constrain young pulsar velocities and proper motions, given  an
association with a supernova remnant.

The standard age estimator for radio pulsars assumes the frequency evolution 
is of the form 
\begin{equation}
\dot\omega = k \omega^n,
\label{eq:torque}
\end{equation}
where $n$ is the ``braking index,'' and $k$ is a constant that depends on 
the magnetic moment of the neutron star.  The braking index can be determined 
from a measurement of the second time derivative of the frequency.  Assuming
$k$ and $n$ to be constant, the age is given by \citep{mt77}:
\begin{equation}
\tau = \frac{P}{(n-1)\dot{P}} \left[1 - \left(\frac{P_0}{P}\right)^{n-1}\right],
\label{eq:spin}
\end{equation}
where $P$ is the rotation period, $\dot{P}$ is its time derivative,
and $P_0$ is the spin period of the pulsar at the time it became a dipole 
rotator, which is generally presumed to coincide with the supernova event. 
The braking index is equal to 3 in a simple vacuum dipole spin-down
model.  For $P_0 << P$ and $n=3$, Equation~1 reduces to 
$\tau_c = P/2\dot{P}$, the often-used pulsar characteristic age.  

For the five pulsars for which a constant value of $n$ has been
measured \citep{lps88,kms+94,lpgc96,dnb99,ckl+00}, the observed
values are in the range 1.4--2.91.  Thus, pulsars do not rotate like
simple vacuum dipoles \citep[e.g.][]{mel97}.  Nevertheless, the range 
of behaviors is limited, and is observationally well constrained.  

The situation is less clear for the initial spin period, $P_0$.  This can
be determined from Equation~1 if the age is known and the braking index
measured. This is only the case for the Crab pulsar, whose estimated
$P_0\sim19$~ms has led to the generally made assumption that the initial
period is much smaller than the current spin period
for all but the very fastest pulsars.  Further support for such short
birth spin periods comes from the existence of a young 16~ms pulsar
PSR~J0537$-$6910 
\citep{mgz+98}.  However, the initial spin period distribution of
neutron stars is not well predicted by theory, since the rotation rates
of the cores of the massive progenitors are largely unknown --
differential rotation could make them deviate significantly from those
of surface layers \citep{es78}.  Furthermore, circumstances at the time
of, or shortly after, core collapse could significantly affect the
neutron star spin independent of the angular momentum properties of the
progenitor \citep[e.g.][]{sp98}.

The supernova remnant G11.2$-$0.3 has received considerable
observational attention because of the possibility that it is
associated with a ``guest star'' witnessed by Chinese astronomers in
the year 386 AD \citep{cs77}.  The remnant's highly circular morphology
and high surface brightness are clear indicators of youth.  \citet{dow84}
postulated that G11.2$-$0.3 is the remnant of a Type II supernova and
suggested that it is a more evolved version of Cas A, the well-known
and much younger oxygen-rich remnant.  
\citet{mr87} reported evidence for a central, flat-spectrum radio
component, suggesting the presence of a neutron star, supporting
the Type II interpretation.  \citet{ggts88} presented a high-resolution
radio map that revealed clumps in the shell, further suggesting that
G11.2$-$0.3 is an evolved Cas A.  In spite of these points,
\citet{rlbs94} argued on the basis of {\it ROSAT} spectral observations
that G11.2$-$0.3 is the remnant of a Type Ia supernova.

The hard X-ray capabilities of the {\it ASCA} satellite were the
key to establishing the true nature of G11.2$-$0.3.
\citet{vad+96}
using {\it ASCA}, reported a hard, non-thermal X-ray
source near the center of the remnant, as would be expected if it
harbored an energetic neutron star, thus confirming the suggestion
made by \citet{mr87}.  However, Vasisht et al. could neither
resolve the nebula, nor determine its precise location within the shell 
because of the $\sim 3'$ spatial resolution of the {\it
ASCA} mirrors.  They did, however, perform nonequilibrium
ionization modeling of the spectrum of the shell.  This put an upper
bound on the age of $\la 2000$~yr, in agreement with the 386 AD
association \citep{aok95}.

Also using {\it ASCA} data, \citet{ttdm97} discovered a 65-ms pulsar
(AX~J1811.5$-$1926) in the direction of G11.2$-$0.3, in agreement with
the hypothesis that the remnant contains an energetic neutron star.
However, surprisingly, \citet{ttd+99} found that it had a
characteristic age of 24,000~yr, in contradiction with the apparent
youth of the remnant, as well as with the tentative 386 AD
association.



One way to verify the association and constrain the age independently
is using high spatial resolution X-ray observations.  
Since pulsars are a high-velocity population, a pulsar located very close
to the geometric center of the remnant implies both a highly probable
association and that the entire system is very young, as insufficient
time must have elapsed for the pulsar to travel far from its birth
place.

High spatial resolution X-ray imaging of young pulsars and supernova
remnants is important for another reason.  Recent {\it Chandra X-ray
Observatory} images of the Crab, Vela and PSR B1509$-$58 pulsars
\citep{wht+00,hgh01,kpg+00} have revealed a wealth of detail regarding
the structures surrounding them, and have argued strongly that pulsar
wind nebulae, also known as ``plerions,'' in general do not not have
simple, spherically symmetric structures as has often been assumed in
models \citep[see][for a review]{got02}.

Here we present {\it Chandra X-ray Observatory} images of G11.2$-$0.3
which reveal, for the first time, the precise projected location of the
pulsar within the shell and the morphology of the pulsar wind nebula.
We restrict our present discussion to the image of the remnant;
detailed results from spectroscopy will be reported separately (Roberts
et al., in preparation).

\section{Observations and Results}

NASA's {\it Chandra X-ray Observatory} observed G11.2$-$0.3 at two
epochs, the first (Sequence Number 50076) on 2000 August 6, and
the second (Sequence Number 50077) on 2000 October 15.  The 
exposure for the first epoch was 20~ks. The second epoch consisted of
two exposures, one of 10~ks and the other of 5~ks.  In all observations,
the remnant was positioned on the back-illuminated CCD chip S3 of
the ACIS instrument in standard exposure mode.  In this mode, the
time resolution (3.2~s) is too coarse to resolve the pulsations from
the pulsar.

The data were analyzed using the CIAO 2.02 and MIRIAD software
packages.  Following the energy binning scheme of \citet{hrbs00}, we
added together the individual count maps from the three different
observing epochs in the 0.6--1.65 keV, 1.65--2.25 keV, and 2.25--7.5
keV energy bands. Spectrally weighted exposure maps were created for
each observation and energy band, and were summed over the three
observations, creating a total count map and exposure map for each
energy band. The count maps were divided by the exposure maps, and the
result convolved with a 5$''$ FWHM Gaussian to enhance the nebular
structure given the low count rate. The three individual maps were then
combined into a 3-color image, with red, green, and blue assigned to
the low, medium, and high energy bands respectively.

In Figure~1 we present the {\it Chandra} image of G11.2$-$0.3.  The
image clearly shows the symmetric, ring-like structure of the shell.
The overall shell X-ray morphology is remarkably like the radio
morphology, with a similar enhancement in the southeast quadrant and
similar clumps along the edge \citep{ggts88}.  At the geometric center
of the shell is a bright point source, the pulsar.  Note that because
of the smoothing, the point source has been broadened in this image
beyond the $\la 1\arcsec$ width of the ACIS point spread function.  
From the unsmoothed image, the
position of the pulsar is (J2000) RA 18$^{\rm h}$~11$^{\rm m}$~29$^{\rm
s}$.22, DEC $-19^{\circ} \; 25' \; 27.''6$.  As the pulsar is a bright
point source, the uncertainty in the position is completely dominated
by the uncertainty in the {\it Chandra} aspect solution.
The nominal 1$\sigma$ radius uncertainty of the latter is 
$0.''6$.\footnote{Chandra Proposers' Observatory Guide, Rev. 3.0, p.
60}  We verified this by optically identifying two sources in the
field-of-view and comparing their positions as reported by {\it
Chandra} with their catalogued positions.  One source (at {\it
Chandra}-reported (J2000) RA 18$^{\rm h}$~11$^{\rm m}$~39$^{\rm s}$.77,
DEC $-19^{\circ} \; 22' \; 5''.1$) was identified in the 2MASS, DSS,
USNO and GSC catalogs, which had positions that agreed with each other
and with the {\it Chandra} position to within 0$''$.5.  The other
source (at {\it Chandra}-reported (J2000) RA 18$^{\rm h}$~11$^{\rm
m}$~41$^{\rm s}$.88, DEC $-19^{\circ} \; 28' \; 16.''0$) was identified
in the 2MASS and USNO catalogs, again with positional consistency of
$0.''6$.  Thus, the {\it Chandra} position appears to be reliable to
within the nominally quoted uncertainty of 0$.''$6.  Given its
newly determined position, and because this is
a rotation-powered pulsar, we
rename the source to be \psr\ and refer to it as such hereafter.


The pulsar wind nebula (PWN) morphology is revealed for the first time
in this image.  There are two distinct emission regions within the
shell.  The hard emission (blue in the image) is collimated, with axis
at $\sim$60$^{\circ}$ east of north. To examine this component better,
we made a 4--9 keV image, which is dominated by non-thermal emission,
and smoothed it with a 2$''$ Gaussian to highlight the fine structure
(see Fig.~2).  The PWN extends for a total of $\sim$40$''$ to the northeast
and southwest of the pulsar.
To the southwest, a
significantly brighter plume starts $\sim$5$''$ from the pulsar, peaks at
$\sim$10$''$, extends to $\sim$20$''$, then becomes much fainter. To the
northeast, the emission remains narrowly confined for $\sim$20$''$ before
``bending'' by $\sim 90^\circ$ and fading.  There is a second emission
region also roughly centered on the pulsar with a much softer spectrum
(red in Fig.~1).  Its long axis lies at an angle of $\sim60^{\circ}$
relative to the hard emission.  It extends roughly $2^{\prime}$ and is
much broader in width than is the hard emission.

\section{Discussion}

The {\it Chandra} image clearly reveals that the pulsar is very close
to the geometric center of the shell.  Fitting circles by eye to the
outer through inner portions of the shell suggest the center is
within a few arcseconds of the pulsar.  In order to verify this
less subjectively, we made a smoothed image of the remnant in which all
pixels had a value of zero below a threshold $\sim 10\sigma$ above the
average background pixel, or unity otherwise, i.e. if they were clearly
part of the remnant.  We then calculated average radial profiles centered at
various points near the center, on a 2.$''$5 grid.
If the shell were a perfect annulus,
the most central position should correspond to the radial profile which
drops to zero at the smallest radius. Since the shell has structure, we
chose 0.01 as our average pixel value cutoff for the shell edge. The
nominal best fit center from this process is to the
southeast of the pulsar (J2000 RA 18$^{\rm h}$~11$^{\rm m}$~29$^{\rm s}$.3, 
DEC $-19^{\circ} \; 25' \; 31.''1$), although only 
profiles from points greater than $\sim
8^{\prime \prime}$ from the pulsar were unambiguously broader than the one
centered on the pulsar. We varied both the image threshold and the edge
cutoff values, and obtained similar results.  We therefore consider
$8^{\prime \prime}$ to be a firm upper limit on the displacement of the
pulsar from the shell center.  This number is dominated by the
difficulty in determining the center of the shell, not the location of
the pulsar.  We also examined the radio image of the remnant, and
visual fitting of circles came up with a best fit center within a few
arcseconds of the pulsar. The broadening of the radio emission in the
southwest quadrant of the remnant causes on/off threshold images to
differ significantly from a circle, making a radial profile
analysis similar to that used for the X-ray data of questionable use.

This result strongly supports the association of the pulsar with the
shell.  Furthermore, the fact that the pulsar is so close to the center
of the remnant provides strong independent evidence for the youth of
the entire system, i.e. that the pulsar characteristic age is an
overestimate of the true age.  This is because, as we now show, if the
system really were 24,000~yr old, the pulsar would probably have moved
significantly away from its birthplace.

First, we discuss distance estimates to the supernova remnant.
These have been made using
neutral hydrogen absorption spectra. \citet{rgmb72} found
$d>5$~kpc.  \citet{bmd85}, upon redoing this observation at the VLA,
found weak evidence for absorption at negative velocities, which they
interpreted as suggestive of a distance of 26~kpc, on the outskirts of
the Galaxy, and implying a diameter of some 30~pc.  This latter estimate is
not supported by any other observation, particularly not the other
indicators of youth, namely the high radio surface brightness and
symmetric structure.  Indeed, as argued by Green et al. (1988), the
lack of any absorption between $+45$~km~s$^{-1}$ and the tangent point
at $+120$~km~s$^{-1}$ argues against a distance outside the solar
circle. Rather, it implies a distance of $\sim$5~kpc, the near distance
corresponding to $+45$~km~s$^{-1}$. The weak absorption at negative
velocities in the spectrum of \citet{bmd85} is probably due
to unusual motions in local gas.  A
distance of 5~kpc implies a diameter of $\sim$6~pc, in agreement with
the apparent youth as inferred from its morphology and surface
brightness, and has been adopted in subsequent studies of the
remnant. The equivalent neutral hydrogen absorption toward the system
as inferred from X-ray observations, $2 \times 10^{22}$~cm$^{-2}$
(Roberts et al., in preparation) is not inconsistent with a distance 
as small as 5~kpc.
No radio pulsations have been detected from the pulsar \citep{ckm+98},
so a dispersion-measure-based distance estimate is not available.

The angular displacement of the pulsar from its birth place would be
 \begin{equation}
\theta = 24'' \left(\frac{v_t}{345 \; {\rm km} \; {{\rm s}^{-1}} }\right) \left(\frac{\tau}{1614 \; {\rm yr}} \right) \left(\frac{d}{5 \; {\rm kpc}} \right)^{-1},
\label{eq:angle}
\end{equation}
where $v_t$ is the pulsar transverse velocity, $\tau$ is the true age, and
$d$ is the distance.  Thus, even in the unlikely event that the system
is as far away as 15~kpc, the pulsar would have to have $v_t < 23$~km~s$^{-1}$
if $\tau = 24,000$~yr, given that $\theta < 8''$.  For $d=5$~kpc at this age, 
and given the constraint on $\theta$, $v_t < 8$~km~s$^{-1}$.

\citet{ll94} showed that the mean pulsar
transverse velocity is 345~km~s$^{-1}$, and inferred a mean 3D
velocity of 450~km~s$^{-1}$.  Although other analyses have
attempted to refine this result \citep{lbh97,hp97,cc98}, it is clear
that pulsars are a high velocity population,
with mean 3D velocity in the range 250--450~km~s$^{-1}$.
\citet{cc98} suggest that the velocity distribution
may have two components, one with 3D mean $\sim$700~km~s$^{-1}$, one
with $\sim$175~km~s$^{-1}$, representing 14\% and 86\% of the population,
respectively.  They speculate regarding a third component having
mean $<50$~km~s$^{-1}$, but conclude it could represent at most 5\% of
pulsars.

We can estimate the probability for occurrence of the low velocity
required by our constraint on $\theta$ (Eq.~\ref{eq:angle}) 
by assuming a reasonable model for the pulsar transverse velocity
distribution.  For a Maxwellian distribution of velocities having mean
transverse speed $\overline{v_t} = 345$~km~s$^{-1}$ \citep{ll94}, the
probability of a pulsar having $v_t < 8$~km~s$^{-1}$ is $<0.1\%$.
By contrast, for $d=5$~kpc and $\tau = 1600$~yr, $v_t <
108$~km~s$^{-1}$, still small compared to the average, but having a
more reasonable 7\% probability of occurring.  Although these
probabilities depend on the true pulsar population velocity distribution
(e.g.,
for $\overline{v_t} = 100$~km~s$^{-1}$, the above probabilities are
0.4\% and 60\%, respectively), the overall conclusion appears firm: the
location of the pulsar is at odds with its characteristic age for any
reasonable distance, unless it has an improbably low transverse
velocity or it happens to be representative of a very small ($<5$\%)
subset of the pulsar population that has very low ($<50$~km~s$^{-1}$)
space velocity \citep{cc98}.  Although we cannot unambiguously rule out
these latter possibilities, given the independent evidence for the
youth of the remnant, we conclude that the pulsar characteristic age is
in all likelihood an overestimate of the true age of the system by a
factor of $\sim 12$.

Figure~3 is a plot of true age versus unknown initial spin period
for four different braking indices representing the observed range,
using Equation~\ref{eq:spin} \citep[see also][]{ttd+99}.  
The true age of the system, $\sim$2000~yr, is
indicated by a straight horizontal line near the bottom of the plot.
To have $P_0 \lapp 20$~ms requires $n \gapp 24$, which is unlikely
given the range of observed $n$'s.  Although there have been
claims of $n$'s outside this narrow range
\citep[e.g.][]{gr78,jg99}, in no case is the measured value known to be
constant.  Specifically, for no case in which a measurement of $n>3$
has been claimed has a repeated measurements yielded the same value. 
Such ``variable'' braking indexes are probably due either to
random timing noise or long-term glitch recovery 
\citep{ch80,sl96}.  Thus, Figure~3 clearly indicates that for any
constant braking index that is consistent with those of the five
pulsars for which it has been measured (\S\ref{sec:intro}), 
the initial spin period of the
G11.2$-$0.3 pulsar must be near $\sim $62~ms.  This is significantly
longer than those of the Crab and N157B pulsars.  So large a difference
may be hinting that the true distribution of initial spin periods of
neutron stars is larger yet.

There are in principle alternative explanations for the fact that the
pulsar's characteristic age is much larger than the true age.
The simple spin-down Equation~\ref{eq:spin} might not
hold.  This could be true if, for example, the magnetic moment were  
not constant, that is $k=k(t)$ in Equation~\ref{eq:torque}.  However,
if so, it is not hard to show \citep{br88} that the
magnetic moment would have to be decay on a time scale of $\lapp
2$~kyr.  This is at odds both with the existence of many older pulsars having
comparable or larger $B$ fields, as well as with
models of magnetic field decay in  neutron stars \citep[e.g.][]{gr92}.

Alternatively, it might be noted that very significant deviations from
simple spin-down have been observed for soft gamma repeaters (SGRs)
\citep[e.g.][]{wkv+99} and anomalous X-ray pulsars (AXPs)
\citep[e.g.][]{opmi98,kgc+01}, both of which have been suggested to be
young, isolated neutron stars.  However the same spin-down processes
operating in those sources are unlikely to be relevant to \psr\ as the
latter has spin period, X-ray spectrum, and spin-down power 
characteristic of rotation-powered pulsars.  The absence of
radio pulsations from the source \citep{ckm+98} is not evidence against
this, given the broad range of pulsar radio luminosities \citep{tml93}.
The SGRs and AXPs, by contrast, cannot be powered by rotation and have
been suggested to be powered by their enormous inferred magnetic fields
-- the ``magnetar'' model \citep{td96a}.  In this model, spin-down
anomalies are a direct result of the large magnetic field, inapplicable
to \psr.  

An alternative model, in which SGRs and AXPs are accreting from a disk
of material that fell back onto the neutron star shortly after the
supernova explosion, has recently been invoked for rotation-powered
pulsars \citep{mlr01,mph01}.  In this model a fall-back disk exerts a
significant torque on the neutron star via the propeller mechanism and
results in braking indexes very different from the vacuum dipole value
(though typically less than 3), and hence characteristic ages very
different from true ages.  However, even if such a disk could survive
the pulsar wind (an issue not addressed by existing studies), it would
be very difficult for the propeller mechanism to spin the pulsar down from a
Crab-like initial spin period on a  time scale of $\sim$2~kyr (M.
Lyutikov, personal communication).

Finally, we note that possible contamination of the pulsar's $\dot{P}$
by Doppler shift due to binary motion is unlikely to explain the age
discrepancy because it would have to exactly cancel out a much larger
intrinsic $\dot{P}$.  On the other hand, if the source were accreting
near its equilibrium spin period, the observed $\dot{P}$ would be due
to accretion torque.  However, there is no evidence for accretion from
the observed X-ray emission, nor is there any evidence for a binary
companion.  In particular, from the 2MASS survey, we find an upper
limit $m_J > 16.5$, which, assuming extinction $A_J=3.0$ \citep{zom90},
rules out all stars of spectral type B5 and earlier and all O and B
giant stars.  Thus, a high-mass X-ray binary is all but ruled out,
while a low-mass X-ray binary is unlikely given the association with
the young supernova remnant.  

Thus, the most conservative conclusion is that the initial spin
period of \psr\ is close to its present period.

Of all known pulsars having characteristic ages under 100~kyr, one
quarter have periods under 90~ms.  Discarding the Crab, N157B and PSR
B0540$-$69 pulsars (all of which have very short current spin periods),
if the remainder were born spinning at $\sim$60~ms, all would have true
ages substantially less than their characteristic ages (see Fig.~3).

Even if an alternative spin-down mechanism is at work and the initial
spin period of the pulsar is not long, the large characteristic age for
\psr\ casts doubt on the characteristic ages of other young pulsars.
This has a variety of implications for young pulsar astronomy.
Smaller true ages drive up
required transverse velocities when attempting to associate a remnant
with a pulsar that is not at its center.
Population synthesis studies have generally assumed
that all pulsars have much shorter Crab-like initial spin periods
\citep[e.g.][]{lbdh93,no90,cz98,ch00,mc00}.  This could be problematic
for some of these studies.  For example, longer
initial spin periods imply lower initial spin-down luminosities, which
could affect the observability of the pulsar population at gamma-ray
energies.  In addition, as tests of neutron star cooling models require
good age constraints \citep[see reviews by][]{tsu86,oge95a}, use of the
characteristic age in these applications may be problematic as well.
A factor of ten age error in the age range $10^3-10^4$~yr
could cause confusion in discriminating among non-standard cooling
models \citep{utn94}.

We note, however, that larger initial spin periods do not affect
the age estimates for the bulk of the pulsar population,
unless the initial spin period distribution is much broader.  
The possibility of the existence of a population of pulsars ``injected''
into the population spinning at a few hundred milliseconds has been
considered in the literature \citep[e.g.][]{ec89,no90} though
more recent studies suggest that this does not occur \citep{lbdh93}.
Indeed, for long period pulsars, $\tau_c$ is more likely to be a
significant underestimate of the true age, since the effects of a
braking index less than 3 becomes important \citep{lpgc96,gf00}.

\subsection{The Pulsar Wind Nebula}

The hard emission centered on the pulsar 
is undoubtedly some form of pulsar wind nebula.  Given its elongated
morphology, we speculate that it is either a torus or disk being viewed
edge-on, or else it is collimated jet-like emission.

A hard X-ray torus has been seen around the Crab pulsar, and is likely
the result of an equatorial outflow interacting with the supernova
ejecta.  \citet{hgh01} have suggested such a disk also exists around
the Vela pulsar, although only a portion of it is observed.  
For G11.2$-$0.3, if the hard emission is a disk like that
in the Crab nebula, it is being viewed very close to edge-on, and would
have to have significant asymmetry about the pulsar.  The brightest
feature in the emission is in the southwest ``arm'' some 10$''$ from
the pulsar.  This corresponds to a size 0.24($d$/5~kpc)~pc, nearly
double the size of the ``inner ring'' around the Crab \citep{wht+00}
but comparable to that of the inner region of the main torus.  Indeed
the entire extent of the hard emission matches in scale with the
overall size of the Crab X-ray torus.  If this interpretation is
correct, it would suggest that the pulsar spin axis lies in the plane
of the sky, perpendicular to the hard-spectrum elongation.  However,
given that the spin-down luminosity of \psr\ is over 60 times smaller
than that of the Crab pulsar, if hot gas in the SNR interior is
containing the PWN, the pressure in G11.2$-$0.3 must be $\sim$8 times
lower than in the Crab nebula.  Also, if this
interpretation is correct, there is no evidence for collimated jets as
seen in the Crab pulsar.  As the jets in the Crab are aligned with the
pulsar's space velocity, the lower inferred velocity of \psr\
could be related to the absence of visible jets.


On the other hand, the elongated hard-spectrum emission may itself be
the analogy of the jets in the Crab pulsar, and may directly delineate
the pulsar spin axis.  The asymmetry about the pulsar could be due to
differences in the densities of the stellar ejecta along the jet axes,
or to Doppler boosting, or to some combination of both.  If this
emission does originate from a jet, it is hard to understand the
apparent $\sim 90^{\circ}$ bend in the emission on the south side,
unless there is a significant density enhancement in that direction.
In this interpretation, there is no clear evidence for a hard X-ray
torus, although the fainter emission out to $\sim 40''$ from the pulsar
and the few arcsecond enhancement around the pulsar perpendicular to
the ``jets" (Fig.~2) are both possibilities.

The other emission component within the shell clearly has a
significantly softer spectrum, and could be an enhancement in the shell
seen in projection. However, its apparent symmetry around the pulsar
suggests a possible association. It could represent material heated by
the forward shock from the pulsar wind, analogous to the heating of
material by the spherical supernova blast-wave. If so, the direction of
elongation is likely to delineate either the equatorial plane of the
pulsar or the polar axis.  A detailed spectral analysis of this and the
other components will be presented elsewhere (Roberts et al., in
preparation).

\section{Conclusions}

Thanks to the superb spatial resolution of {\it Chandra}, we have
determined that the 65-ms pulsar in G11.2$-$0.3, which we designate
PSR~J1811$-$1925,  is at the precise geometric center of the remnant,
formally to within $8''$.  This provides strong support for the
pulsar's association with the shell, consistent with the 386 AD guest
star, and in agreement with the shell properties.  However, the
inferred age, $\sim2000$~yr (or 1615~yr if the 386 AD association
holds), is at odds with the much larger characteristic age of the
pulsar.  For the reasonable spin-down assumptions, this suggests that
this pulsar had birth spin period very close to its present spin
period.  This result is insensitive to the pulsar's braking index,
assuming the latter is not very different fromm those measured for
other pulsars.  This result calls into question the reliability of
characteristic ages as true age estimators for a significant fraction
of young pulsars \citep[see also][]{lpgc96,gf00}.

In addition, these {\it Chandra} observations have, for the first time,
revealed the morphology of the pulsar wind nebula at the center of the
supernova remnant.  Its elongated morphology at hard X-ray energies,
like those seen in high resolution X-ray observations of the Crab and
Vela pulsars adds to the growing evidence that such structures are
ubiquitous around young pulsars, and demand explanation.

\section*{Acknowledgements}

We thank David Helfand, Maxim Lyutikov, and Fotis Gavriil 
for useful discussions.  This
work was supported in part by {\it Chandra} grant GO0-1132X from the
Smithsonian Astrophysical Observatory, NASA LTSA grant NAG5-8063, and
NSERC research grant RGPIN228738-00 to VMK and by a Quebec Merit
Fellowship to MSER.


\clearpage
\begin{figure}
\plotone{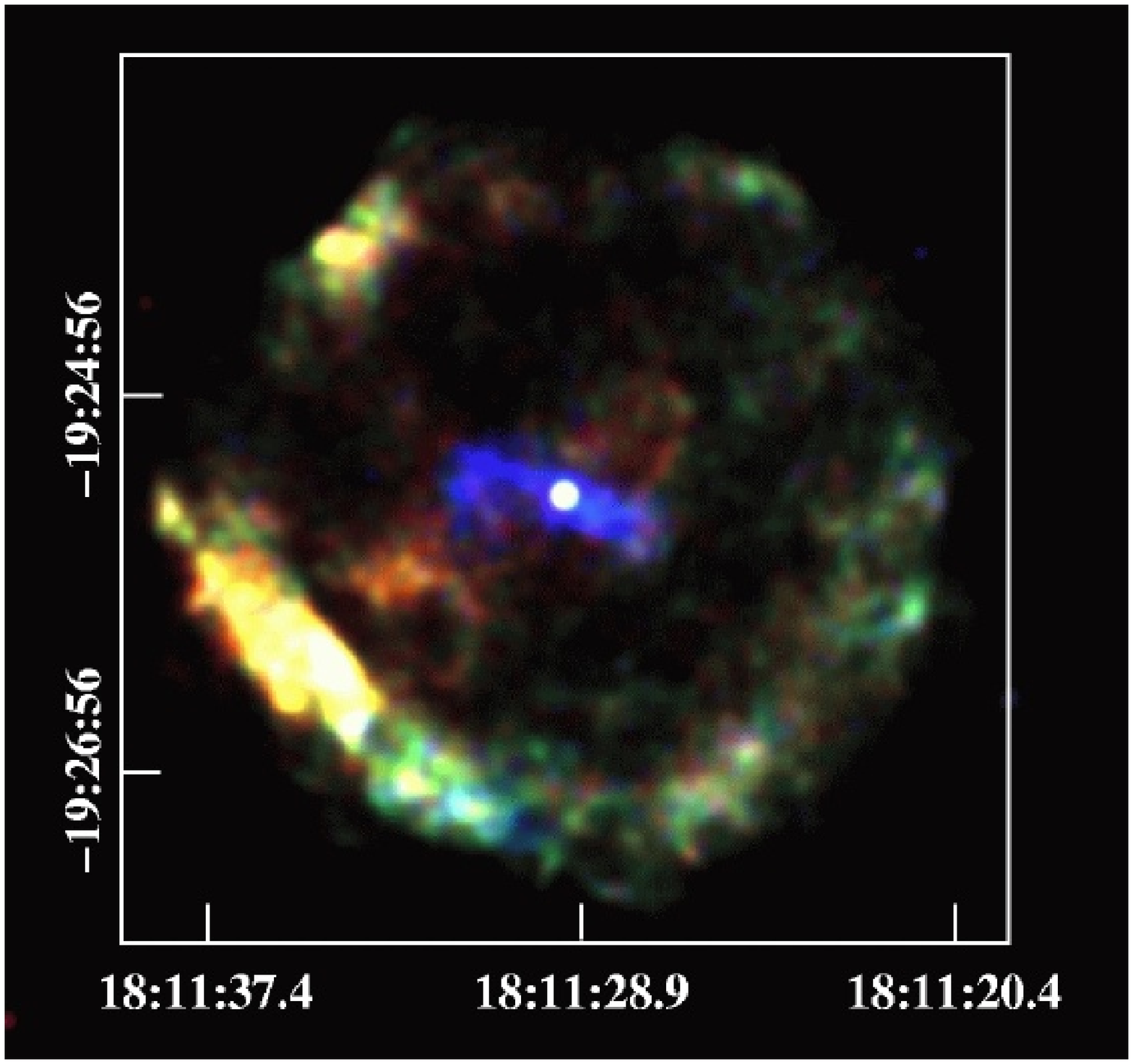}
\figcaption{{\it Chandra} image of G11.2$-$0.3, smoothed by a 5$''$
Gaussian, and color coded as follows:  red represents photons of
energies 0.6--1.65~keV, green is 1.65--2.25~keV, and blue 2.25--7.5~keV.
}
\label{ximage}
\end{figure}

\clearpage
\begin{figure}
\plotone{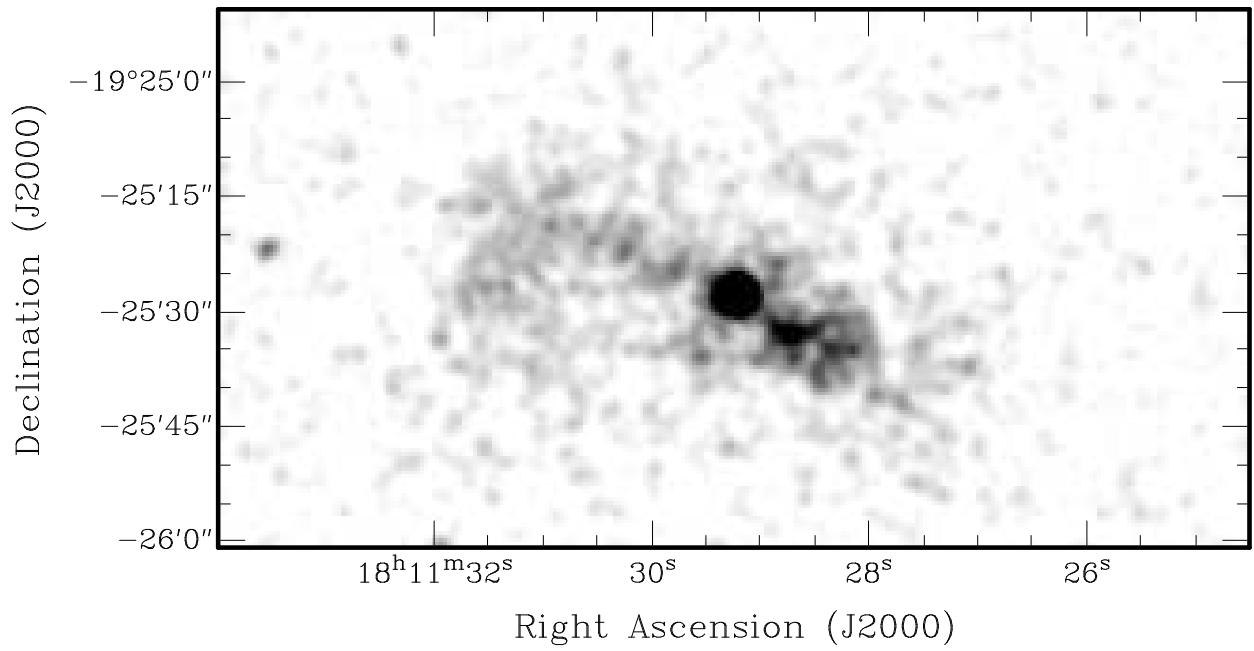}
\figcaption{
The central PWN in hard (4-9~keV) X-rays, smoothed by a 2$''$
Gaussian.}
\label{ximage:inset}
\end{figure}

\clearpage

\begin{figure}
\plotone{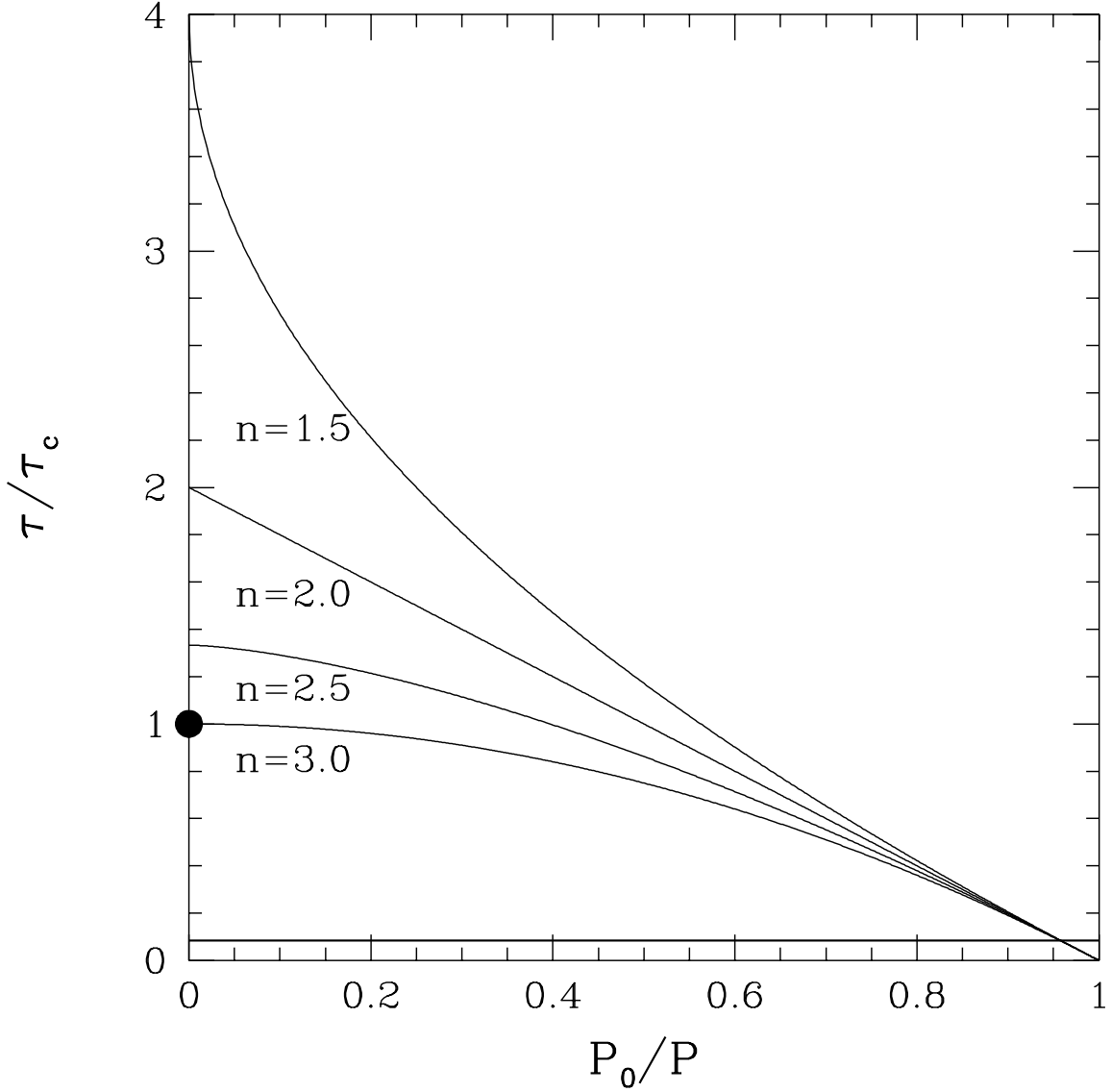}
\figcaption{True age $\tau$, in units of the characterstic age $\tau_c
\equiv P/2\dot{P}$, versus initial spin period $P_0$, in units of the
current spin period $P$, for four different braking indexes $n$.  
For G11.2$-$0.3, $\tau_c = 24,000$~yr, and $P=65$~ms.
The dot corresponds to the conventional assumptions of $n=3$ and $P_0 <<
P$.  The horizontal line indicates the true age of the system as
estimated from the remnant properties as well as the observations
reported here.  It demonstrates that for any braking index within the
observed range, the initial spin period of the pulsar had to be
$\sim$62~ms.} \label{fig:age} \end{figure}

\end{document}